\documentclass{article}
\usepackage{iftex}
\ifTUTeX
  \usepackage{fontspec}
\else
  \usepackage[T1]{fontenc}
  \usepackage[utf8]{inputenc} 
  \DeclareUnicodeCharacter{200B}{{\hskip 0pt}}
\fi
\usepackage{spconf,amsmath,graphicx}
\usepackage{caption}
\usepackage{subcaption}
\usepackage{siunitx}
\usepackage{bm}
\usepackage{amsfonts}
\usepackage{graphicx}
\usepackage{amsmath}
\usepackage{multirow}
\usepackage{textcomp}
\usepackage{tabularx,booktabs}
\usepackage{makecell}
\usepackage{footnote}
\usepackage[hyphens,spaces,obeyspaces]{url}
\usepackage[hidelinks]{hyperref}

\title{StyleMelGAN: An Efficient High-Fidelity Adversarial Vocoder with Temporal Adaptive Normalization}
\name{Ahmed Mustafa \qquad Nicola Pia \qquad Guillaume Fuchs}

\address{Fraunhofer IIS, Erlangen, Germany \\
\texttt{\{ahmed.mustafa.ahmed,nicola.pia,guillaume.fuchs\}@iis.fraunhofer.de}}

\begin{document}
\topmargin=0mm
%
\maketitle
\begin{abstract}
In recent years, neural vocoders have surpassed classical speech generation approaches in naturalness and perceptual quality of the synthesized speech. 
Computationally heavy models like WaveNet and WaveGlow achieve best results, while lightweight GAN models, e.g. MelGAN and Parallel WaveGAN, remain inferior in terms of perceptual quality. 
We therefore propose StyleMelGAN, a lightweight neural vocoder allowing synthesis of high-fidelity speech with low computational complexity. 
StyleMelGAN employs temporal adaptive normalization to style a low-dimensional noise vector with the acoustic features of the target speech. 
For efficient training, multiple random-window discriminators adversarially evaluate the speech signal analyzed by a filter bank, with regularization provided by a multi-scale spectral reconstruction loss. 
The highly parallelizable speech generation is several times faster than real-time on CPUs and GPUs. 
MUSHRA and P.800 listening tests show that StyleMelGAN outperforms prior neural vocoders in copy-synthesis and Text-to-Speech scenarios.
\end{abstract}
\begin{keywords}
Neural Vocoder, GANs, Neural PQMF, Speech Synthesis, TADE
\end{keywords}
%

\section{Introduction}
\label{SEC_intro}


Neural vocoders have proven to outperform classical approaches in the synthesis of natural high-quality speech in many applications, such as text-to-speech, speech coding, and speech enhancement.
The first groundbreaking generative neural network to synthesize high-quality speech was WaveNet~\cite{wavenet}, and shortly thereafter many other approaches were developed~\cite{waveglow,samplernn,wavernn}.
These models offer state-of-the-art quality, but often at a very high computational cost and very slow synthesis.
An abundance of models generating speech with lowered computational cost was presented in the recent years. 
Some of these are optimized versions of existing models~\cite{parallelwavenet}, while others leverage the integration with classical methods~\cite{lpcnet}.
Besides, many completely new approaches were also introduced, often relying on GANs~\cite{melgan,pwgan,gantts}.
Most GAN vocoders offer very fast generation on GPUs, but at the cost of compromising the quality of the synthesized speech.

The main objective of this work is to propose a GAN architecture, which we call StyleMelGAN, that can synthesize very high quality speech at low computational cost and fast training.
StyleMelGAN's generator network contains \SI{3.86}{M} trainable parameters, and synthesizes speech at \SI{22.05}{\kHz}​ around \SI{2.6}{x} faster than real-time on CPU and more than \SI{129}{x} on GPU.
The synthesis is conditioned on the mel-spectrogram of the target speech, which is inserted in every generator block via Temporal Adaptive DE-normalization (TADE), a feature modulation firstly introduced in image synthesis~\cite{spade}.
This approach for inserting the conditioning features is very efficient and, as far as we know, new in the audio domain.
The adversarial loss is computed by an ensemble of four discriminators, each operating after a differentiable Pseudo Quadrature Mirror Filter bank (PQMF).
This permits to analyze different frequency bands of the speech signal during training.
In order to make the training more robust and favor generalization, the four discriminators are not conditioned on the input acoustic features used by the generator, and the speech signal is sampled using random windows as in~\cite{gantts}.

To summarize our contributions:
\begin{itemize}
	\itemsep0em 
	\item We introduce StyleMelGAN, a low complexity GAN vocoder for high-fidelity speech synthesis conditioned on mel-spectrograms via TADE layers. The generator is highly parallelizable and completely convolutional.
	\item The aforementioned generator is trained adversarially with an ensemble of PQMF multi-sampling random window discriminators regularized by a multi-scale spectral reconstruction loss.
	\item We assess the quality of the generated speech using both objective (Fr\' echet scores) and subjective metrics. To this end we present the results of two listening tests, a MUSHRA test for the copy-synthesis scenario and a P.800 ACR test for the TTS one, both confirming that StyleMelGan achieves state-of-art speech quality.
\end{itemize}


\section{Related works}
\label{SEC_related_works}


Existing neural vocoders usually synthesize speech signals directly in time-domain, by modelling the amplitude of the final waveform.
Most of these models are generative neural networks, i.e. they model the probability distribution of the speech samples observed in natural speech signals.
They can be divided in autoregressive, which explicitely factorize the distribution into a product of conditional ones, and non-autoregressive or parallel, which instead model the joint distribution directly.
Autoregressive models like WaveNet~\cite{wavenet}, SampleRNN~\cite{samplernn} and WaveRNN~\cite{wavernn} have been reported to synthesize speech signals of high perceptual quality~\cite{prachi}.
A big family of non-autoregressive models is the one of Normalizing Flows, e.g. WaveGlow~\cite{waveglow}. 
A hybrid approach is the use of Inverse Autoregressive Flows~\cite{parallelwavenet}, which use a factorized transformation between a noise latent representation and the target speech distribution.

Early applications of GANs~\cite{gans} for audio include WaveGAN~\cite{wavegan} for unconditioned audio generation, and GanSynth~\cite{gansynth} for music generation.
MelGAN~\cite{melgan} learns a mapping between the mel-spectrogram of speech segments and their corresponding waveforms.
It ensures faster than real-time generation and leverages adversarial training of multi-scale discriminators regularized by a feature matching loss.
GAN-TTS~\cite{gantts} is the first GAN vocoder to use uniquely adversarial training for speech generation conditioned on linguistic features.
Its adversarial loss is calculated by an ensemble of conditional and unconditional random window discriminators.
Parallel WaveGAN~\cite{pwgan} uses a generator, similar to WaveNet in structure, trained using an unconditioned discriminator regularized by a multi-scale spectral reconstruction loss.
Similar ideas are used in Multiband-MelGAN~\cite{mbmelgan}, which generates each sub-band of the target speech separately, saving computational power, and then obtains the final waveform using a synthesis PQMF.
Research in this field is very active and we can cite the very recent GAN vocoders such as VocGan~\cite{vocgan} and HiFi-GAN~\cite{kong2020hifi}.

\section{StyleMelGAN}
\label{SEC_stylemelgan}

\subsection{Generator architecture}

\begin{figure}[htb]
\begin{minipage}[b]{1.0\linewidth}
  \centering
  \centerline{\includegraphics[width=\linewidth]{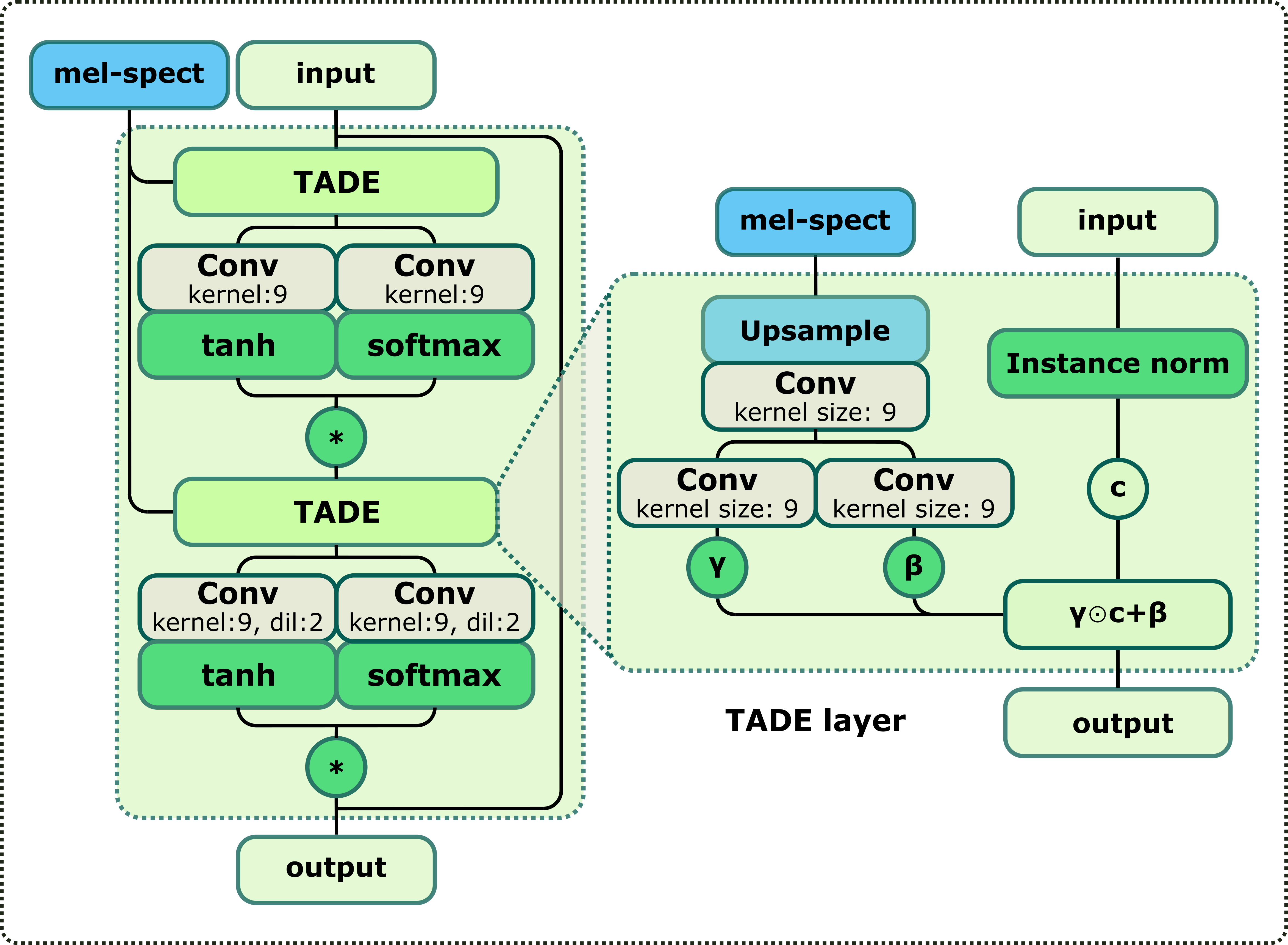}}
\end{minipage}
\caption{The TADEResBlock and TADE layer.}
\label{IMG_tade}
\end{figure}

The generator model maps a noise vector $z \sim \mathcal{N}(0,I_{128})$ into a speech waveform at \SI{22.05}{\kHz} by progressive upsampling conditioned on mel-spectrograms.
It uses Temporal Adaptive DE-normalization (TADE), which is a feature-wise conditioning based on linear modulation of normalized activation maps.
The modulation parameters $\gamma$ and $\beta$ are adaptively learned from the conditioning features, and have the same dimension as the input activation signal.
This technique was first used for semantic image synthesis in~\cite{spade}, where the modulation parameters are learned in spatial domain (SPADE).
This delivers the conditioning features to all layers of the generator model and hence preserving the signal structure at all upsampling stages.
Figure~\ref{IMG_tade} illustrates the structure of the TADE block.
The input activation is firstly normalized via instance normalization~\cite{ulyanov2016instance} to create a content feature map $c$ that is adaptively modulated via $\gamma\odot c+\beta$, where $\odot$ indicates element-wise multiplication. The combination of normalization and modulation represents a style transfer process, which is the central component of the generator model.

\begin{figure}[htb]
\begin{minipage}[b]{1.0\linewidth}
  \centering
  \centerline{\includegraphics[width=1.0\linewidth]{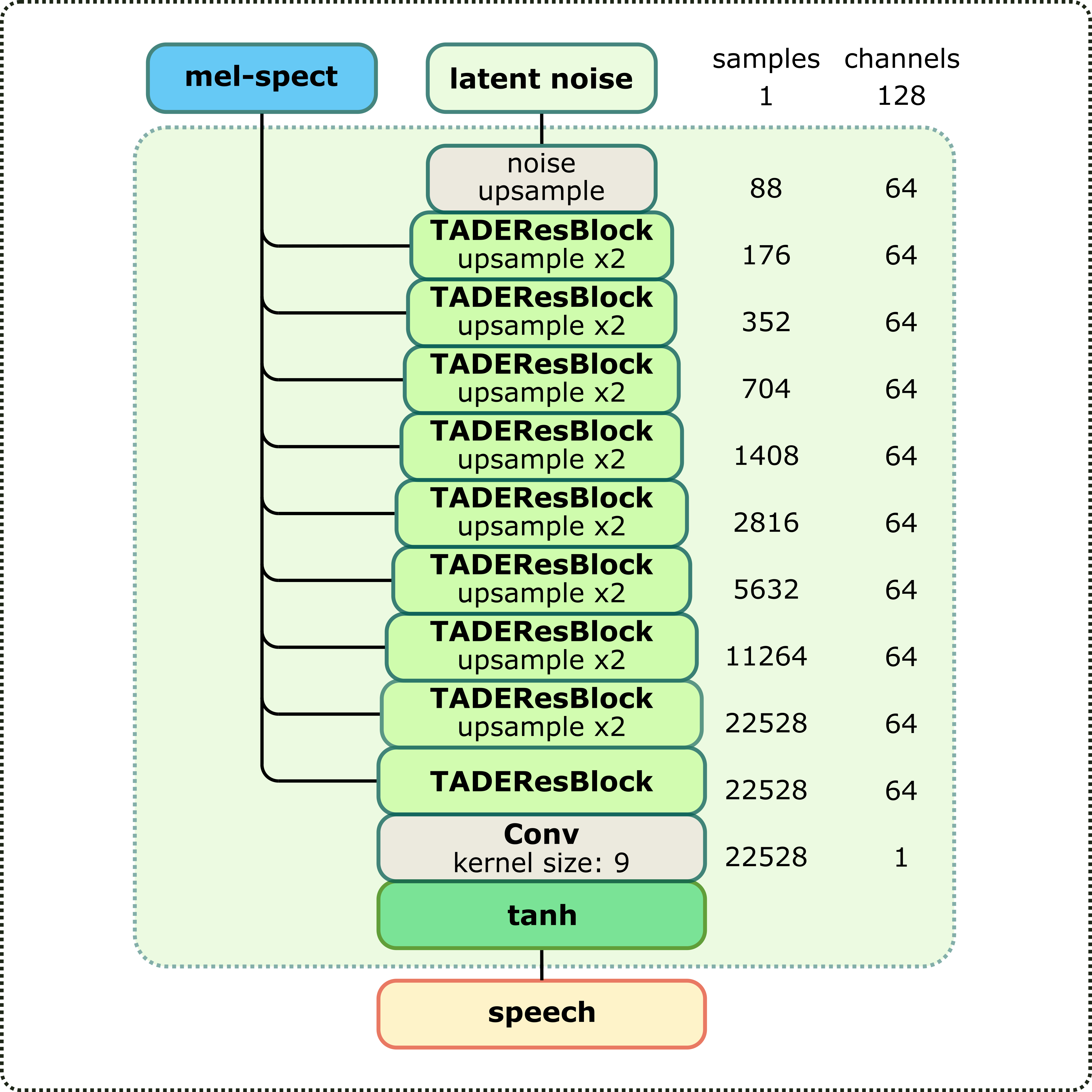}}
\end{minipage}
\caption{StyleMelGAN generator architecture.}
\label{IMG_g}
\end{figure}

We use softmax-gated $\tanh$ activation functions, which reportedly performs better than ReLU.
Softmax gating was proposed in~\cite{Mustafa2019} and allows for less artifacts in audio waveform generation.
The TADEResBlock in Figure~\ref{IMG_tade} is the basic building block of the generator model.
The complete generator architecture is shown in Figure~\ref{IMG_g}.
It includes eight upsampling stages, each consisting of a TADEResBlock and a layer upsampling the signal by a factor two, plus one final activation module.
The final activation comprises one TADEResBlock followed by a channel-change convolutional layer with $\tanh$ non-linearity.
This design allows to use a channel depth of just $64$ for the convolution operations with maximum dilation factor of $2$, thus reducing computational complexity.

\subsection{Filter Bank Random Window Discriminators}

\begin{figure}[htb]
\begin{minipage}[b]{1.0\linewidth}
  \centering
  \centerline{\includegraphics[width=1.0\linewidth]{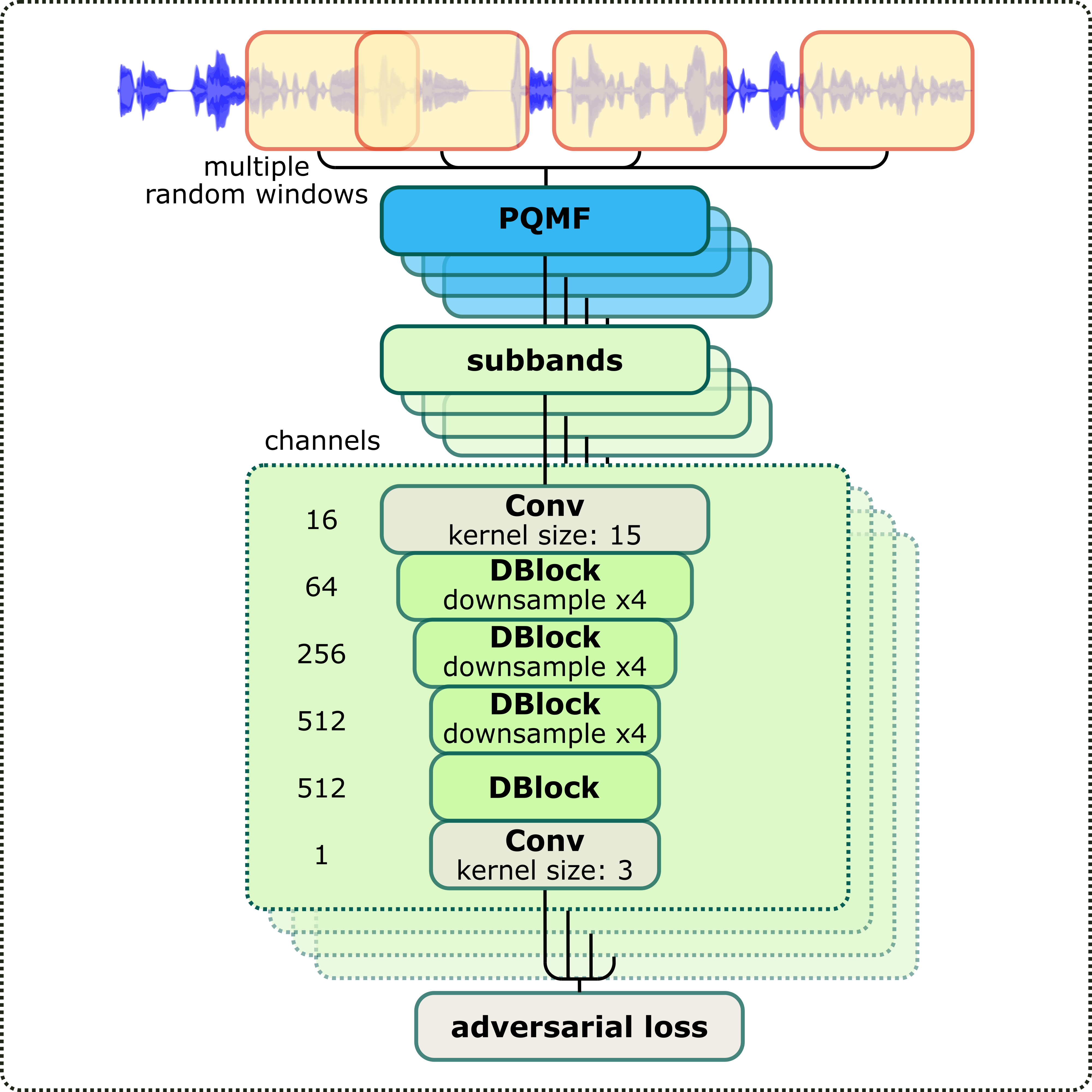}}
\end{minipage}
\caption{StyleMelGAN discriminator architecture. Each DBlock consists of a 1D-convolution layer followed by a LeakyReLU as in~\cite{mbmelgan}}
\label{IMG_d}
\end{figure}

StyleMelGAN uses four discriminators for its adversarial training, each based on the architecture proposed in~\cite{mbmelgan}, but without average pooling for downsampling.
Moreover, each discriminator operates on a random window sliced from the input speech waveform as proposed in~\cite{gantts}. 
Finally, each discriminator analyzes the sub-bands of the input speech signal obtained by an analysis PQMF~\cite{nguyen1994near}.
More precisely, we use $1$, $2$, $4$, and $8$ sub-bands calculated respectively from random windows of $512$, $1024$, $2048$, and $4096$ samples extracted from a waveform of one second.
Figure~\ref{IMG_d} illustrates the complete architecture of the proposed filter bank random window discriminators (FB-RWDs). 
This enables a multi-resolution adversarial evaluation of the speech signal in both time and frequency domains.

\subsection{Training procedure}
Training GANs from scratch is known to be challenging.
Using random initialization of the weights, the adversarial loss can lead to severe audio artifacts and unstable training.  
To avoid this problem, we follow the same training procedure used in~\cite{pwgan}.
At first, the generator is pretrained using only the spectral reconstruction loss $\mathcal{L}_\textup{aux}$ consisting of error estimates of the spectral convergence and the log-magnitude computed from different STFT analyses defined by Equation (6) in~\cite{pwgan}.
The generator obtained in this fashion is able to generate very tonal signals with significant smearing at high frequencies.
This is nonetheless a good starting point for the adversarial training, which can benefit from a better harmonic structure than if it had been started directly from a complete random noise signal.
The adversarial training then drives the generation to naturalness by removing the tonal effects and sharpening the smeared frequency bands.
We use the hinge loss to evaluate the adversarial metric of each discriminator $D_{k}$ with the same formulation given by Equation (1) in~\cite{melgan} but for $k = 1,2,3,4$.
The spectral reconstruction loss $\mathcal{L}_\textup{aux}$ is still used for regularization to prevent the emergence of adversarial artifacts.
Hence, the final generator objective is:
\begin{equation}
\min_{G}\Big(\mathbb{E}_{z}\Big[\sum_{k=1}^4 -D_{k}\left(G(s,z)\right)\Big]+\mathcal{L}_\textup{aux}(G)\Big),
\end{equation}
where $s$ represents the conditioning features (e.g., mel-spectrograms) and $z$ is the latent noise.
Weight normalization~\cite{salimans2016weight} is applied to all convolution operations in $G$ and $D_{k}$.

\section{Experiments}


\subsection{Experimental Setup}

\label{SEC_exp}
In our experiments, we train StyleMelGAN using one NVIDIA Tesla V100 GPU on the LJSpeech corpus~\cite{ljspeech17} at \SI{22.05}{\kHz}.
We calculate the log-magnitude mel-spectrograms with the same hyper-parameters and normalization described in~\cite{pwgan}.
We pretrain the generator for \SI{100}{k} steps using the Adam optimizer~\cite{kingma2014adam} with learning rate $lr_g=10^{-4}$, $\beta = \{0.5,0.9\}$. 
When starting the adversarial training, we set $lr_g=5*10^{-5}$ and use FB-RWDs with the Adam optimizer of $lr_d=2*10^{-4}$, and same $\beta$. 
The FB-RWDs repeat the random windowing for multiple times at every training step to support the model with enough gradient updates. 
We use a batch size of $32$ with segments of length $1$ second for each sample in the batch.
The training lasts for about \SI{1.5}{M} steps.

\subsection{Evaluation}
\label{SEC_quality}

\begin{table}[th]
\caption{cFDSD scores for different neural vocoders (lower is better).}
\label{cfdsd}
\vspace{-4mm}
\begin{center}
\scalebox{0.9}{
\begin{tabular}{ c|c|c|c } 

\hline
\textbf{Model} & \textbf{Source} & \textbf{Train cFDSD} & \textbf{Test cFDSD} \\
\hline
 MelGAN & \cite{melgan} & 0.235 & 0.227 \\ 
 P. WaveGAN & \cite{hayashi2020espnet}& 0.122 & 0.101\\ 
 WaveGlow & \cite{prachi} & 0.099 & 0.078 \\ 
 WaveNet & \cite{hayashi2020espnet} & 0.176 & 0.140 \\ 
 StyleMelGAN & ours & \textbf{0.044} & \textbf{0.068} \\ 
\hline
\end{tabular}}
\end{center}
\end{table}

We perform objective and subjective evaluations of StyleMelGAN against other neural vocoders trained on the same dataset.
For all our evaluations, the test set is composed of unseen items recorded by the same speaker and randomly selected from the LibriVox online corpus\footnote{\url{https://librivox.org/reader/11049}}.

Traditional objective measures such as PESQ and POLQA are not reliable to evaluate speech waveforms generated by neural vocoders~\cite{prachi}.
Instead, we use the conditional Fr\' echet Deep Speech Distances (cFDSD) defined in~\cite{gantts} and implemented in~\cite{gritsenko2020spectral}. 
Table~\ref{cfdsd} shows that StyleMelGAN outperforms other adversarial and non-adversarial vocoders.

\begin{figure}[htb]
\begin{minipage}[b]{1.0\linewidth}
  \centering
  \centerline{\includegraphics[width=1.0\linewidth]{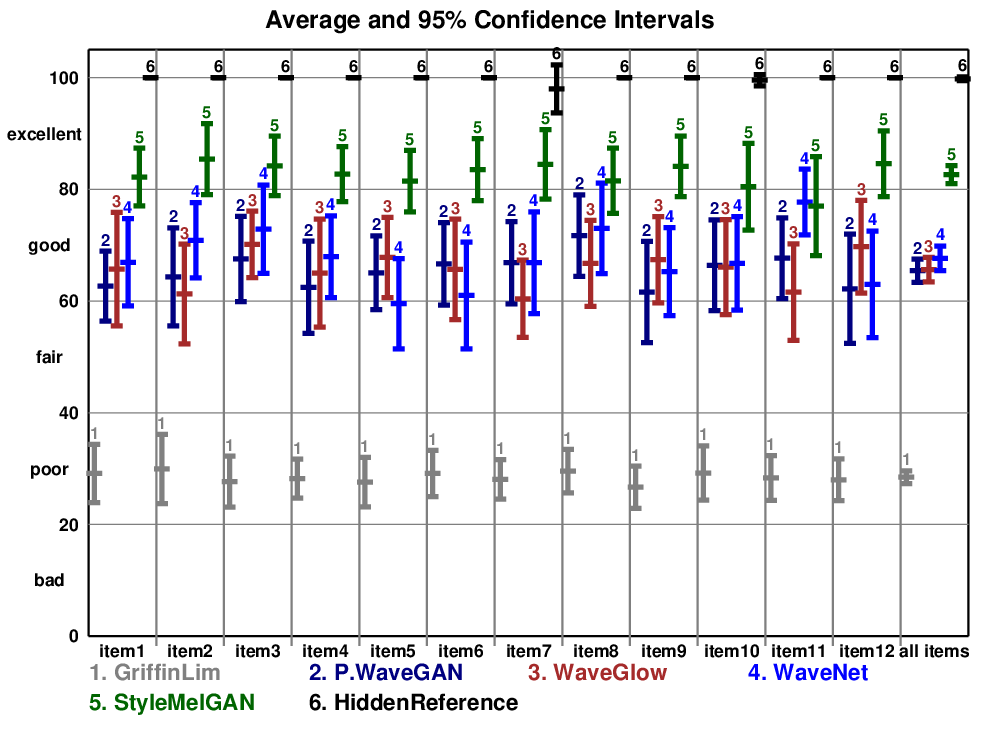}}
\end{minipage}
\caption{The results of our MUSHRA expert listening test.}
\label{IMG_lt}
\end{figure}

For copy-synthesis, we conducted a MUSHRA listening test with a group of $15$ expert listeners, whose results are shown in Figure~\ref{IMG_lt}.
We chose this type of test as in~\cite{lpcnet,prachi}, because this permits to more precisely evaluate the generated speech.
The anchor is generated using the PyTorch implementation of the Griffin-Lim algorithm with $32$ iterations. 
StyleMelGAN significantly outperforms the other vocoders by about $15$ MUSHRA points.  
The results also show that WaveGlow produces outputs of comparable quality to WaveNet, as already found in~\cite{prachi}, while being on par with Parallel WaveGAN (P.WaveGAN).

For Text-to-Speech, we evaluate the subjective quality of the audio outputs via a P.800 ACR~\cite{p800} listening test performed by 37 listeners in a controlled environment. 
The Transformer.v3 model from~\cite{hayashi2020espnet} was used to generate mel-spectrograms of the test set.
We also add a Griffin-Lim anchor as replacement of MNRU~\cite{p800} to calibrate the judgement scale.
Table~\ref{TBL_P800} shows that StyleMelGAN clearly outperforms the other models.

\begin{table}[th]
\caption{Results of P.800 ACR listening test.}
\label{TBL_P800}
\vspace{-4mm}
\begin{center}
\begin{tabular}{ c|c} 
\hline
\textbf{System} & \textbf{MOS} \\
\hline
 GriffinLim & $1.30 \pm 0.03$ \\ 
 Transformer + P. WaveGAN & $3.17 \pm 0.06$ \\  
 Transformer + WaveNet & $3.82 \pm 0.06$  \\ 
 Transformer + StyleMelGAN & $\bm{4.00 \pm 0.06}$  \\ 
 Recording & $4.28 \pm 0.05$  \\
\hline
\end{tabular}
\end{center}
\end{table}

We report the computational complexity in table~\ref{TBL_complexity}, which shows the generation speed in real-time factor (RTF) and the number of parameters for different parallel vocoder models.
StyleMelGAN provides a clear compromise between quality and inference speed\footnote{Check our demo samples at the following url: \url{https://fhgspco.github.io/smgan/}}.

\begin{table}[th]
\caption{Number of parameters and real-time factors for generation on CPU (Intel Core i7-6700 3.40 GHz) and GPU (Nvidia Tesla V100 GPU) for various models under study.}
\label{TBL_complexity}
\vspace{-4mm}
\begin{center}
\begin{tabular}{ c|c|c|c } 

\hline
\textbf{Model} & \textbf{Parameters} & \textbf{CPU} & \textbf{GPU} \\
\hline
 MelGAN & \SI{4.26}{M} & \SI{7}{x} & \SI{286}{x} \\ 
 P. WaveGAN & \SI{1.44}{M} & \SI{0.8}{x} & \SI{38}{x} \\  
 StyleMelGAN & \SI{3.86}{M} & \SI{2.6}{x} & \SI{129}{x} \\ 
 WaveGlow & \SI{80}{M} & - & \SI{10}{x} \\ 
\hline
\end{tabular}
\end{center}
\end{table}

\section{Conclusion}
\label{SEC_conclusion}
This work presents StyleMelGAN, a light-weight and efficient adversarial vocoder for high-fidelity speech synthesis faster than real-time on both CPUs and GPUs. 
The model uses temporal adaptive normalization (TADE) to condition each layer of the generator on the mel-spectrogram of the target speech.
For adversarial training, the generator competes against Filter Bank Random Window Discriminators that analyze the synthesized speech signal at different time scales and multiple frequency sub-bands. 
Both objective evaluations and subjective listening tests show that StyleMelGAN significantly outperforms prior adversarial and maximum-likelihood vocoders, providing a new state-of-the-art baseline for neural waveform generation. 
Possible future prospects include further complexity reductions to support applications running on low-power processors. 

\section{Acknowledgments}
\label{SEC_ack}
The authors would like to thank Christian Dittmar, Prachi Govalkar, Yigitcan Özer, Srikanth Korse, Jan Büthe, Kishan Gupta and Markus Multrus from Fraunhofer IIS for their devoted support and helpful tips.

\bibliographystyle{IEEEbib}
\bibliography{smgan}

\begin{thebibliography}{10}

\bibitem{wavenet}
A.~van~den Oord, S.~Dieleman, H.~Zen, K.~Simonyan, et~al.,
\newblock ``{WaveNet: A Generative Model for Raw Audio},''
\newblock {\em arXiv:1609.03499}, 2016.

\bibitem{waveglow}
R.~Prenger, R.~Valle, and B.~Catanzaro,
\newblock ``{Waveglow: A Flow-based Generative Network for Speech Synthesis},''
\newblock in {\em IEEE International Conference on Acoustics, Speech and Signal
  Processing (ICASSP)}, 2019, pp. 3617--3621.

\bibitem{samplernn}
S.~Mehri, K.~Kumar, I.~Gulrajani, R.~Kumar, et~al.,
\newblock ``{SampleRNN: An Unconditional End-to-End Neural Audio Generation
  Model},''
\newblock {\em arXiv:1612.07837}, 2016.

\bibitem{wavernn}
N.~Kalchbrenner, E.~Elsen, K.~Simonyan, S.~Noury, et~al.,
\newblock ``Efficient neural audio synthesis,''
\newblock {\em arXiv:1802.08435}, 2018.

\bibitem{parallelwavenet}
A.~van~den Oord, Y.~Li, I.~Babuschkin, K.~Simonyan, et~al.,
\newblock ``{Parallel WaveNet: Fast High-Fidelity Speech Synthesis},''
\newblock in {\em Proceedings of the 35th ICML}, 2018, pp. 3918--3926.

\bibitem{lpcnet}
J.~Valin and J.~Skoglund,
\newblock ``{LPCNET: Improving Neural Speech Synthesis through Linear
  Prediction},''
\newblock in {\em IEEE International Conference on Acoustics, Speech and Signal
  Processing (ICASSP)}, 2019, pp. 5891--5895.

\bibitem{melgan}
K.~Kumar, R.~Kumar, de~T.~Boissiere, L.~Gestin, et~al.,
\newblock ``{MelGAN: Generative Adversarial Networks for Conditional Waveform
  Synthesis},''
\newblock in {\em Advances in NeurIPS 32}, pp. 14910--14921. 2019.

\bibitem{pwgan}
R.~Yamamoto, E.~Song, and J.~Kim,
\newblock ``{Parallel Wavegan: A Fast Waveform Generation Model Based on
  Generative Adversarial Networks with Multi-Resolution Spectrogram},''
\newblock in {\em IEEE International Conference on Acoustics, Speech and Signal
  Processing (ICASSP)}, 2020, pp. 6199--6203.

\bibitem{gantts}
M.~Bi\'nkowski, J.~Donahue, S.~Dieleman, A.~Clark, et~al.,
\newblock ``{High Fidelity Speech Synthesis with Adversarial Networks},''
\newblock {\em arXiv:1909.11646}, 2019.

\bibitem{spade}
T.~Park, M.~Y. Liu, T.~C. Wang, and J.~Y. Zhu,
\newblock ``{Semantic Image Synthesis With Spatially-Adaptive Normalization},''
\newblock in {\em Proc. of the IEEE Conference on Computer Vision and Pattern
  Recognition (CVPR)}, 2019.

\bibitem{prachi}
P.~Govalkar, J.~Fischer, F.~Zalkow, and C.~Dittmar,
\newblock ``{A Comparison of Recent Neural Vocoders for Speech Signal
  Reconstruction},''
\newblock in {\em Proceedings of the {ISCA} Speech Synthesis Workshop}, 2019,
  pp. 7--12.

\bibitem{gans}
I.~Goodfellow, J.~Pouget-Abadie, M.~Mirza, B.~Xu, et~al.,
\newblock ``{Generative Adversarial Nets},''
\newblock in {\em Advances in NeurIPS 27}, pp. 2672--2680. 2014.

\bibitem{wavegan}
C.~Donahue, J.~McAuley, and M.~Puckette,
\newblock ``{Adversarial Audio Synthesis},''
\newblock {\em arXiv:1802.04208}, 2018.

\bibitem{gansynth}
J.~Engel, K.~K. Agrawal, S.~Chen, I.~Gulrajani, et~al.,
\newblock ``{GANSynth: Adversarial Neural Audio Synthesis},''
\newblock {\em arXiv:1902.08710}, 2019.

\bibitem{mbmelgan}
G.~Yang, S.~Yang, K.~Liu, P.~Fang, et~al.,
\newblock ``{Multi-band MelGAN: Faster Waveform Generation for High-Quality
  Text-to-Speech},''
\newblock {\em arXiv:2005.05106}, 2020.

\bibitem{vocgan}
J.~Yang, J.~Lee, Y.~Kim, H.~Cho, and I.~Kim,
\newblock ``{VocGAN: A High-Fidelity Real-time Vocoder with a
  Hierarchically-nested Adversarial Network},''
\newblock {\em arXiv:2007.15256}, 2020.

\bibitem{kong2020hifi}
Jungil Kong, Jaehyeon Kim, and Jaekyoung Bae,
\newblock ``Hifi-gan: Generative adversarial networks for efficient and high
  fidelity speech synthesis,''
\newblock {\em arXiv preprint arXiv:2010.05646}, 2020.

\bibitem{ulyanov2016instance}
D.~Ulyanov, A.~Vedaldi, and V.~Lempitsky,
\newblock ``{Instance normalization: The missing ingredient for fast
  stylization},''
\newblock {\em arXiv:1607.08022}, 2016.

\bibitem{Mustafa2019}
A.~Mustafa, A.~Biswas, C.~Bergler, J.~Schottenhamml, and A.~Maier,
\newblock ``{Analysis by Adversarial Synthesis - A Novel Approach for Speech
  Vocoding},''
\newblock in {\em Proc. Interspeech}, 2019, pp. 191--195.

\bibitem{nguyen1994near}
T.~Q. Nguyen,
\newblock ``{Near-perfect-reconstruction pseudo-QMF banks},''
\newblock {\em IEEE Transactions on Signal Processing}, vol. 42, no. 1, pp.
  65--76, 1994.

\bibitem{salimans2016weight}
T.~Salimans and D.~P. Kingma,
\newblock ``{Weight normalization: A simple reparameterization to accelerate
  training of deep neural networks},''
\newblock in {\em Advances in NeurIPS}, 2016, pp. 901--909.

\bibitem{ljspeech17}
K.~Ito and L.~Johnson,
\newblock ``{The LJ Speech Dataset},''
  {\small\url{https://keithito.com/LJ-Speech-Dataset/}}, 2017.

\bibitem{kingma2014adam}
D.~P. Kingma and J.~Ba,
\newblock ``{Adam: A method for stochastic optimization},''
\newblock {\em arXiv:1412.6980}, 2014.

\bibitem{hayashi2020espnet}
T.~Hayashi, R.~Yamamoto, K.~Inoue, T.~Yoshimura, et~al.,
\newblock ``{Espnet-tts: Unified, reproducible, and integratable open source
  end-to-end text-to-speech toolkit},''
\newblock in {\em IEEE International Conference on Acoustics, Speech and Signal
  Processing (ICASSP)}. IEEE, 2020, pp. 7654--7658.

\bibitem{gritsenko2020spectral}
A.~Gritsenko, T.~Salimans, R.~van~den Berg, J.~Snoek, and N.~Kalchbrenner,
\newblock ``{A Spectral Energy Distance for Parallel Speech Synthesis},''
\newblock {\em arXiv:2008.01160}, 2020.

\bibitem{p800}
``{P.800 : Methods for subjective determination of transmission quality},''
\newblock Standard, International Telecommunication Union, 1996.

\end{thebibliography}

\end{document}